# Text/Graphics Separation and Skew Correction of Text Regions of Business Card Images for Mobile Devices

Ayatullah Faruk Mollah, Subhadip Basu, and Mita Nasipuri

**Abstract**—Separation of the text regions from background texture and graphics is an important step of any optical character recognition system for the images containing both texts and graphics. In this paper, we have presented a novel text/graphics separation technique and a method for skew correction of text regions extracted from business card images captured with a cell-phone camera. At first, the background is eliminated at a coarse level based on intensity variance. This makes the foreground components distinct from each other. Then the non-text components are removed using various characteristic features of text and graphics. Finally, the text regions are skew corrected for further processing. Experimenting with business card images of various resolutions, we have found an optimum performance of 98.25% (recall) with 0.75 MP images, that takes 0.17 seconds processing time and 1.1 MB peak memory on a moderately powerful computer (DualCore 1.73 GHz Processor, 1 GB RAM, 1 MB L2 Cache). The developed technique is computationally efficient and consumes low memory so as to be applicable on mobile devices.

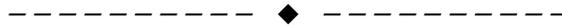

## 1 INTRODUCTION

PERVASIVE availability of mobile phones with built-in cameras has drawn the attention of a number of researchers towards developing camera based applications for handheld mobile devices. Automatic Business Card Reader (BCR) is an example of such applications. Using this application, necessary contact information can be directly populated into the phonebook from the business cards. Although, such applications are commercially available in some mobile handsets (Motorola MOTOROKR E6, Sony Ericsson P1i, etc.) and some softwares such as Abbyy Business Card Reader that is installable in camera phones, the accuracy is yet to be improved to incorporate more complex cards and make such applications really useful in real life scenarios.

The key challenges in the development of a BCR on a mobile phone are the following.

**Deformations in acquired images.** Unlike scanned images, camera captured images suffer from a number of distortions and degradation. As the business cards are planar objects, the acquired images suffer from both skew and perspective distortion. Skew occurs when the camera axis is not properly aligned with that of the object and perspective distortion occurs when a business card is not parallel to the imaging plane. As both these adjustments are manually performed, perfection is quite difficult and so such distortion is very likely to happen, however small be it, for all camera captured business card images.

Quality of the captured images is greatly affected due to improper lighting condition of the surroundings. In absence of sufficient light, one may opt for using camera flash. But, in that case, the centre of the card image becomes brightest and the intensity decays outward [1]. The acquired images may be blurred in case of improper focusing and if either the card or the camera is not perfectly static at the time of capturing.

**Diversity in nature of business cards.** Text of a business card may be cursive, underlined, artistic, and of various alignment, fonts and sizes. Artistic impression, logo, graphics and pictures are often found in both foreground and background of the image. Many a times, they are not distinct from each other and merged with texts and themselves in an overlapping or blowing fashion. Business cards may contain dark text over light background, light text over dark background or both of these.

**Computational constraints of the mobile devices.** Computation under mobile devices suffers from three major computing constraints viz. poor computing power (200 MHz - DualCore 333 MHz ARM Processors), low working memory (22-128 MB RAM) and no *Floating Point Unit* (FPU) for floating point arithmetic operations. Algorithms designed for mobile devices should avoid floating point arithmetic. Otherwise, such arithmetic computations have to be converted to integer ones, that

────────────────

- *A. F. Mollah is with the School of Mobile Computing and Communication, Jadavpur University, Kolkata-700032, India.*
- *S. Basu is with the Department of Computer Science and Engineering, Jadavpur University, Kolkata-700032, India.*
- *M. Nasipuri is with the Department of Computer Science and Engineering, Jadavpur University, Kolkata-700032, India.*

again increases the computational time.

In order to recognize the text information from the card, the text regions and background graphics must be separated. Isolation of text regions from business cards of complex nature is an active area of research. For practical applications, computational time and/or memory intensive algorithms can not be directly implemented on mobile devices in spite of their good performances. For this purpose, a computationally efficient yet powerful text separation method is designed in our work towards developing an efficient BCR for mobile devices.

## 2 LITERATURE SURVEY

Most of the text/graphics separation techniques found in the literature are implemented and evaluated on document images with desktop computers. Very few references are there on text/graphic separation for camera captured documents on mobile devices. However, the following references [2-8] may be cited in this regard.

Jang et al. [2] have classified the equally partitioned 8x8 blocks as information blocks and background blocks based on low frequency DCT coefficients and information pixel density. The information blocks are further classified into character blocks or picture blocks using a similar approach. As the technique removes the non-textual blocks, it can be considered as a text/graphics separation method. Performance of this technique is limited to low resolution images such as VGA resolution. The text/graphics separation method in the work of Shin et al. [3] applies a modified quadratic filter to enhance the blocks classified as character block using low frequency DCT coefficients. Character blocks are binarized and the remaining blocks are removed. The performance is greatly deteriorated if the images are not ill-conditioned e.g. poor resolution, weak illumination, shadow, etc.

Guo et al. [4] have presented a text/image separation algorithm aimed at low resource consumption to run on mobile device. Pilu et al. [5] in their work on light weight text image processing for handheld embedded cameras, proposed a text detection method that sometimes fails to remove the logo(s) of a card and the technique often mistakes parts of the oversized fonts as background and can not deal with reverse text i.e. light texts on dark background. Some other text extraction methods are reported in [6-8].

In [6], text lines are extracted from Chinese business card images using document geometrical layout analysis method. Fisher's Discrimination Rate (FDR) based approach followed by various text selection rules is presented in place of mathematical morphology based operations in [7].

Yamaguchi et al. [8] have designed a digit extraction method for their work on telephone number identification and recognition from signboards. Roberts filter has been used to detect edges from the initial images. Then non-digit components are checked out according to some criteria based on various properties of digits. Hough transform has been used for skew and slant angle estimation.

Apart from the research works cites in the literature [2-8] several commercial applications are also available for the said purpose. As discussed earlier, Motorola, Sony Ericsson, and several other mobile handsets recently came with preinstalled BCR software in mobile phones. Also Abbyy's BCR can be downloaded from [9] and installed into some specific camera phones. However, performance of these commercial systems is limited to simple business cards with minimal graphic contents. Many such systems work only with predefined camera alignments and fair lighting conditions.

The objective of the current work is to develop an improved algorithm for text/graphic separation and skew correction of text regions extracted from business card images for mobile devices, an important preprocessing step to actual BCR system that can compete with the available state-of-the-art in the said domain.

## 3 PRESENT WORK

The current work discussed in this paper may be subdivided into three key modules, viz. background elimination from the camera captured business card images, graphics separation and skew estimation for each text regions of the image. These are discussed in the following subsections.

### 3.1 Background Elimination

At first, the camera captured business card images are virtually split into small blocks. A block is part of either background or a foreground component. This classification is done on the basis of intensity variation within the block. The intensity variation is defined as the difference between the maximum ($G_{max}$) and the minimum ($G_{min}$) gray scale intensity within the block. It is observed that the intensity variation of a text/foreground block is considerably more than that of a background block. A block is classified as a foreground block if following two conditions are satisfied.

Firstly, the intensity range of a block must fall between a heuristically chosen threshold ($\lambda$) and the maximum intensity i.e. 255. Secondly, the intensity variation of the block must be less than an adaptive threshold ($T_\sigma$) as formulated in Eq. 1-2. Otherwise, it is considered as part of a foreground component. It may be noted that not only the intensity range but also the position of the intensity range in the whole intensity band i.e. 0-255 matters in the present technique. Based on where the intensity range lies in the whole intensity band, $T_\sigma$ adapts for a better classification.

The threshold $T_\sigma$ has two components, i.e. a fixed one ($T_{fixed}$) and a variable one ($T_{var}$). $T_{fixed}$ is a constant subject to tuning and $T_{var}$ is formulated in Eq. 2. It is observed from the equation that $G_{min}$ must be greater than $\lambda$. This reveals the reality that even if the intensity variation of a block is less than $T_\sigma$, it is not classified as background until the minimum intensity of the block exceeds $\lambda$. It reduces the possibility of misclassification of foreground blocks as background ones.

$$T_\sigma = T_{fixed} + T_{var} \quad (1)$$

$$T_{var} = [(G_{min}-\lambda) - min(T_{fixed}, G_{min}-\lambda)] * 2 \quad (2)$$

It is evident from Eq. 2 that the computation of $T_\sigma$ is such that the more is the average intensity within the block, the larger is the threshold. In other words, if the intensity band of the block falls towards the higher range of the overall intensity band, then $T_\sigma$ becomes larger. Such formulation helps to efficiently eliminate the background blocks from the captured business card images. Also light backgrounds get easily eliminated in the said approach. The business card image shown in Fig.1(a) looks like Fig.1(b) after applying this technique.

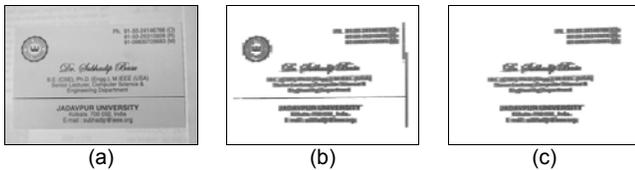

Fig. 1. Different transformation on a sample business card image. (a) A business card image with graphic components, (b) Foreground components are isolated, (c) Graphic components are removed.

### 3.2 Graphics Separation

The four-connected region growing algorithm [10] is applied to identify the distinct foreground Connected Components (CC) from background eliminated card images. A CC may be a picture, logo, texture, graphics, noise or a text region. In the current work, we focus to identify only the text regions using rule-based classification technique. The following features are used to classify a CC under consideration as a text region or not. The technique is described in detail in [11-12]

The height, width, width to height ratio (aspect ratio), gray pixel density, black pixel density, number of vertical and horizontal segments, and the number of cuts (i.e. changes in thresholded intensity divided by two) along the middle row of the CC are considered as features to decide upon the characteristic of each CCs. Different heuristically chosen adaptive (with respect to the size/resolution of the input image) thresholds are estimated for designing the rule-based classifier for text/graphics separation. Too small regions that are unlikely to become text regions and horizontal/vertical lines detected by checking their width, height and aspect ratio are considered as non-text components. Typically, a text region has a certain range of width to height ratio ($R_{w2h}$). So, we consider a CC as a potential text region if $R_{w2h}$ lies within the range ($R_{min}$, $R_{max}$). We assume that neither horizontal nor vertical lines can be drawn through a logo and it is larger than the largest possible character within the card. Thus, logos and other components satisfying the above specification get eliminated. Another important property of text regions is that the number of foreground pixels in a text region is significantly less than that of the background pixels. We consider a certain range of ratio of the foreground pixels to the background ($RA_{cc}$) given by ($RA_{min}$, $RA_{max}$) for the candidate text regions. The graphic components of Fig.1(b) are removed as shown in Fig.1(c).

### 3.3 Skew Angle Estimation and Correction

Skewness in camera captured business card images may appear due to two primary reasons, firstly, misalignment of the handheld mobile camera with respect to the horizontal axis during image capturing, and secondly, due to perspective distortion within the captured image. In the later case, different text regions of any card image may be aligned at different skew angles with respect to the horizontal axis. To address this problem, in the present work, skew angles are estimated and subsequently corrected for each of the connected text regions.

To calculate the skew angle, we consider the bottom profile of a text region. Texts are surrounded by gray rectangular blocks found in Section 3.1. As a card of perfect white background is captured through a camera, the background of the acquired image does not become perfectly white. It becomes a little gray. So, the blocks around a text look gray. The profile contains the heights in terms of pixels from the bottom edge of the bounding rectangle of the text region to the first gray pixel found while moving upward. These heights are measured along the width of the text region. However, if the extent of the gray shade along a column of the profile is too small, we discard it as an invalid profile.

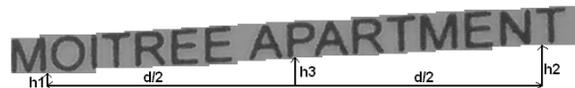

Fig. 2. Skew angle computation for a typical text region

The bottom profile of each text region is represented as an array of heights ($h[i]$s) from the bottom edge of the bounding rectangle, as shown in Fig. 2. Let the width of the text region be $N$ pixels. So, the mean ($\mu$) and the mean deviation ($\tau$) of the profile elements are calculated as given in Eq. 3 and 4 respectively. Then we exclude the profiles that are not within the range (+$\tau$, -$\tau$). Among the rest of the heights, the leftmost ($h1$), rightmost ($h2$) and middle ($h3$) heights are taken into consideration for skew angle computation. Thus, we find $\alpha$, $\beta$ and $\gamma$ as the skew angles obtained from the pairs $h1$-$h2$, $h1$-$h3$ and $h3$-$h2$ respectively. The mean of them is considered as the computed skew angle of the text region. Results of the skew correction method are shown in Fig. 3 for some sample text regions extracted from different business card images.

$$\mu = \frac{1}{N}\sum_{i=0}^{N-1} h[i] \quad (3)$$

$$\tau = \frac{1}{N}\sum_{i=0}^{N-1} |\mu - h[i]| \quad (4)$$

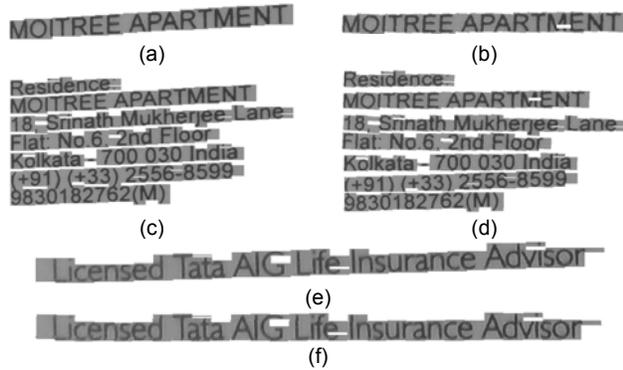

Fig. 3. Sample text regions are shown before and after skew correction. (a) Single line connected component, (b) Skew corrected look of (a), (c) A text region containing multiple lines, (d) Deskewed text region for (c), (e) Large single text line, (f) View of (e) after skew correction

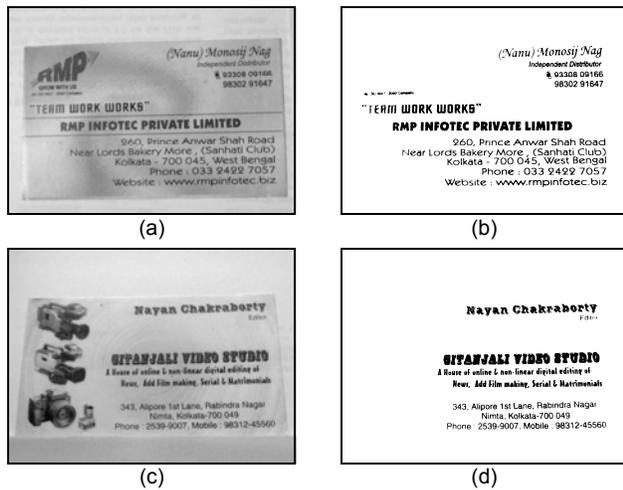

Fig. 4. Sample card images and their graphics eliminated snapshots. (a) A business card image with textual logo and multi-colored background, (b) Extracted texts from the card image (a), (c) A business card image containing multiple pictures in the background, (d) Graphics eliminated view of card image (c)

## 4 EXPERIMENTAL RESULTS AND DISCUSSION

To evaluate the performance of the present technique for text/graphics separation, experiments have been carried out on a dataset of 100 business card images of various types, acquired with a cell-phone camera (Sony Ericsson K810i). The dataset contains both simple and complex cards including complex backgrounds and logos. Some cards contain multiple logos and some logos are combination of text and graphics. Most of the images are skewed, perspectively distorted and degraded. Sample images from business card dataset are shown in Fig. 4(a), 4(c), 5(a), 5(c) and 9(a). A subset of the business card dataset used in the present work is available at [13] for academic/non-commercial purposes.

### 4.1 Performance

Ground truth image is compared with the resultant image for evaluating the performance of the present technique. A component may be either a text or a graphic component. Here, a graphic component refers to all non-text regions including background texture and noises. Based on the presence of a component in either or both the ground truth image and the resultant image, we count the number of true positive ($C_{TP}$), false positive ($C_{FP}$), true negative ($C_{TN}$) and false negative ($C_{FN}$) as shown in Table 1.

TABLE 1
COMPARISON BETWEEN
GROUND TRUTH AND RESULTANT IMAGE

| Ground Truth \ Resultant Image | Text | Graphic |
|---|---|---|
| Text | TP | FN |
| Graphic | FP | TN |

The recall ($R$), precision ($P$) and accuracy ($A$) rates are calculated as formulated in Eq. 5-7. The recall parameter signifies how many text components have been correctly identified among all the text components in the ground truth image whereas the precision factor signifies how many text components identified in the resultant image are truly text components. In an ideal situation $R$, $P$ and $A$ should be all 100 %.

$$R = \frac{C_{TP}}{C_{TP} + C_{FN}} \quad (5)$$

$$P = \frac{C_{TP}}{C_{TP} + C_{FP}} \quad (6)$$

$$A = \frac{C_{TP} + C_{TN}}{C_{TP} + C_{FP} + C_{TN} + C_{FN}} \quad (7)$$

Experiments have been conducted with images of various resolutions of the same set of business cards with $T_{fixed}$ = 20, $\lambda$ = 100, $R_{min}$ = 1.2, $R_{max}$ = 32, $RA_{min}$ = 5 and $RA_{max}$ = 90. The mean $R$, $P$ and $A$ as obtained with various resolutions are shown in Table 2. Fig. 4-5 shows some examples where the technique successfully identifies the text regions of the business card images. The *Receiver Operating Characteristic* (ROC) curve for the present technique is shown in Fig. 6, which is basically a graphical plot of *False Positive Rate* (FPR) vs. *True Positive Rate* (TPR) for all business card images of 3 mega-pixel resolution. The graph shows the effectiveness of the technique.

TABLE 2
CLASSIFICATION ACCURACY OF BUSINESS CARD IMAGE WITH VARIOUS RESOLUTIONS

| Resolution | Recall | Precision | Accuracy |
|---|---|---|---|
| 640x480 (0.3 MP) | 98.07 | 97.21 | 96.69 |
| 800x600 (0.45 MP) | 98.40 | 94.59 | 96.00 |
| **1024x768 (0.75 MP)** | **98.25** | **96.77** | **97.38** |
| 1182x886 (1 MP) | 98.35 | 95.29 | 96.66 |
| 1672x1254 (2 MP) | 98.23 | 96.60 | 97.66 |
| 2048x1536 (3 MP) | 98.96 | 97.21 | 98.00 |

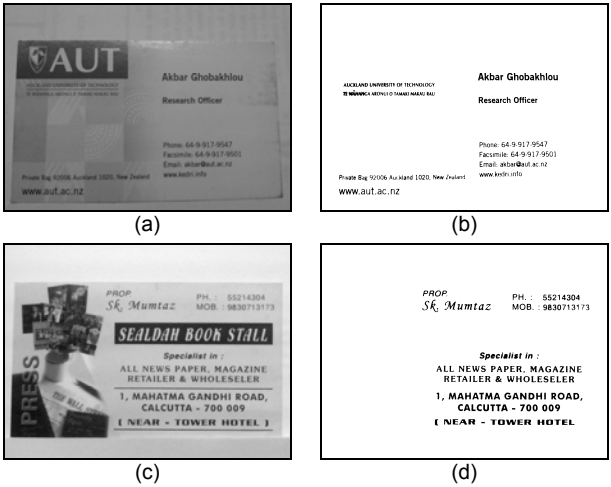

Fig. 5. Successful text/graphics separation for business card images. (a) A dark business card image with unevenly patterned background, (b) Background and non-text components eliminated view of card image (a), (c) A bright business card containing picture and a light text on dark background, (d) Dark texts on light background are extracted from card (c)

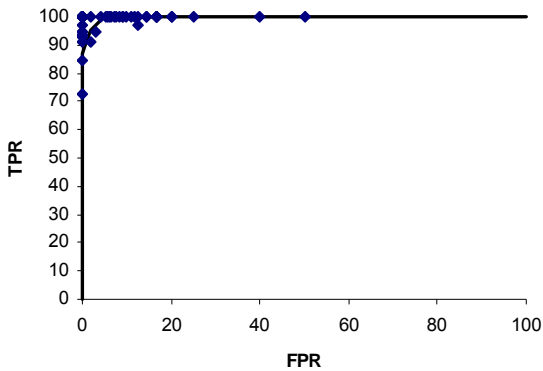

Fig. 6. ROC curve for 3 mega-pixel business card images

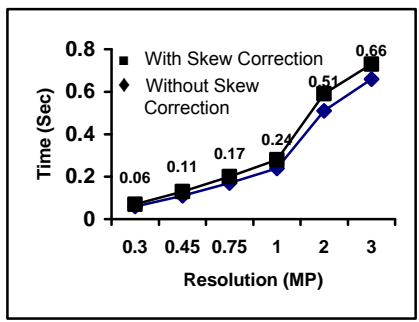

Fig. 7. Time Computation with Various Image Resolutions

### 4.2 Computational Requirements

The applicability of the presented technique on mobile devices is evaluated by its computational requirements. It may be noted that experiments have been conducted on a moderately powerful desktop (DualCore 1.73 GHz Processor, 1 GB RAM, 1 MB L2 Cache). Although, our ultimate objective is to deploy the proposed method into mobile devices, we tried to develop a light-weight Business Card Reader (BCR) system on a desktop to get a first hand idea about the timing and memory requirements for the proposed technique. This information will be helpful in deciding the applicability of the technique in mobile devices.

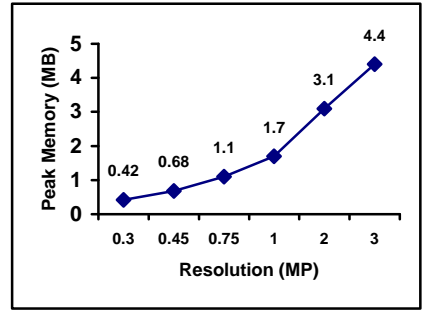

Fig. 8. Memory Consumption with Various Image Resolutions

An observation [14] reveals that the majority of the processing time of an Optical Character Recognition (OCR) engine, embedded into a mobile device, is consumed in preprocessing including skew correction and binarization. Although, we have shown the computational time of the presented method with respect to a desktop, the total time required to run the developed method on mobile devices will be tolerable. Fig. 7-8 show the timing and memory requirements of the present technique with various resolutions respectively. It may be observed from Fig. 7 that time requirements for skew correction technique, an essential component for the overall efficiency of the BCR, is minimal in comparison to the time required for text/graphics separation and binarization modules combined.

### 4.3 Discussion

Although, one can observe that the developed method works well as shown in Fig. 4, Fig. 5 and Table 2, it too has certain limitations. Sometimes, the dot of 'i' or 'j' gets removed during CC analysis. When text and image/logo are very close to each other, they together often form a single CC and get wrongly classified as background or text as shown in Fig. 9. Also, textual graphics (e.g. logo of different companies) get easily classified as potential text regions. The current technique is not applicable for white texts on a dark background as shown in Fig. 5(c-d). These areas may however be considered as our future directions of research in this field. However, text regions are less likely to be classified as graphics as the heuristics are chosen in a conservative way and the graphics identified as texts have the possibility to get removed in subsequent steps. So, by the present technique, we have the least possibility to loose any textual information in the BCR system. On the other hand, the graphics elements that are classified as texts may be subsequently eliminated in further steps like segmentation, recognition and post-processing. These steps are yet to be incorporated in our continuing work.

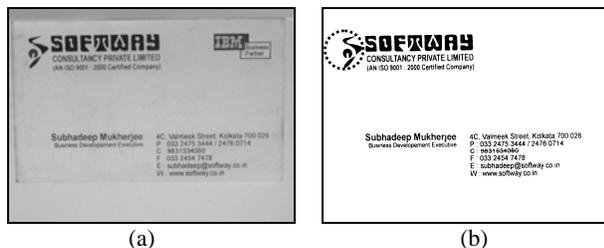

Fig. 9. Sample business card images with partial misclassification, (a) An input image with graphics close a text region, (b) Failed to remove graphics components encircled with dotted lines

## 5 CONCLUSION

In the current work, we have developed a fairly accurate text/graphics separation methodology for camera captured business card images. The present work also implements a fast skew correction technique for the misaligned text regions and subsequent binarization of the gray scale background/graphics suppresses images. Although, the computational requirements have been shown with respect to a PC, the same may be acceptable for mobile architectures (200 MHz - DualCore 333 MHz ARM Processors, 22-128 MB RAM) as well. Observation reveals that with the increase in image resolution, the computation time and memory requirements increase proportionately. It is evident from Fig. 7 that the computationally efficient skew correction algorithm contributes to an average of 15% rise in the overall computation time. Although, the maximum accuracy is obtained with 3 mega pixel resolution, it involves comparatively high memory requirement and 0.66 seconds of processing time. It is evident from the findings that the optimum performance is achieved at 1024x768 (0.75 MP) pixels resolution with a reasonable accuracy of 98.25% (recall) and significantly low (in comparison to 3 MP) processing time of 0.17 seconds and memory requirement of 1.1 MB. In brief, the current work may be viewed as a significant step towards development of effective segmentation/recognition technique for designing a complete and efficient BCR system for mobile devices.

**Acknowledgments.** Authors are thankful to the *Center for Microprocessor Application for Training Education and Research* (CMATER) and project on *Storage Retrieval and Understanding of Video for Multimedia* (SRUVM) of the Department of Computer Science and Engineering, Jadavpur University for providing infrastructural support for the research work. We are also thankful to the *School of Mobile Computing and Communication* (SMCC) for proving the research fellowship to the first author.

## REFERENCES

[1] F. Fisher, "Digital camera for document acquisition", Proc. of the symposium on document image understanding technology, pp. 75–83, 2001

[2] I. H. Jang, C. H. Kim and N. C. Kim, "Region Analysis of Business Card Images in PDA Using DCT and Information Pixel Density", ACIVS 2005, pp. 243-251.

[3] K. T. Shin, I. H. Jang and N. C. Kim, "Block adaptive binarization of ill-conditioned business card images acquired in a PDA using a modified quadratic filter", IET Image Processing, Vol. 1, No. 1, March 2007, pp. 56-66.

[4] J. K. Guo and M. Y. Ma, "A Low Resource Consumption Image Region Extraction Algorithm for Mobile Devices", ICME 2007, pp. 336-339.

[5] M. Pilu and S. Pollard, "A light-weight text image processing method for handheld embedded cameras", BMVC, 2002, (http://www.hpl.hp.com/personal/mp/docs/ bmvc2002-textpipeline.pdf)

[6] W. Pan, J. Jin, G. Shi and Q. R. Wang, "A System for Automatic Chinese Business Card Recognition", ICDAR, 2001, pp. 577-581.

[7] N. Ezaki, K. Kiyota, B. T. Minh, M. Bulacu and L. Schomaker, "Improved Text-Detection Methods for a Camera-based Text Reading System for Blind Persons" , ICDAR, 2005, pp. 257-261.

[8] T. Yamaguchi, Y. Nakano, M. Maruyama, H. Miyao and T. Hananoi, "Digit Classification on Signboards for Telephone Number Recognition", ICDAR, 2003, pp. 359-363.

[9] ABBYY Business Card Reader, http://www.abbyy.com/bcr

[10] E. Gose, R. Johnsonbaugh and S. Jost, "Pattern Recognition and Image Analysis", Prentice-Hall of India, Eastern Economy Edition, pp. 334.

[11] A. F. Mollah, S. Basu, N. Das, R. Sarkar, M. Nasipuri and M. Kundu, "Binarizing Business Card Images for Mobile Devices", accepted for presentation in the Second International Conference on Advances in Computer Vision and Information Technology, Dec 2009.

[12] A. F. Mollah, S. Basu, M. Nasipuri and D. K. Basu, "Text/Graphics Separation for Business Card Images for Mobile Devices", Proceedings of the Eighth IAPR International Workshop on Graphics Recognition (GREC`09), July, 2009, France, pp. 263-270.

[13] http://www.cmaterju.org/research/datasets/businesscards/set1.zip

[14] M. Laine and O. S. Nevalainen, "A Standalone OCR System for Mobile Cameraphones", 17th Annual IEEE International Symposium on Personal, Indoor and Mobile Radio Communications, Sept. 2006, pp. 1-5.

**Ayatullah Faruk Mollah** received his B.E. degree in Computer Science and Engineering from Jadavpur University, Kolkata, India in 2006. Then he served as a Senior Software Engineer in Atrenta (I) Pvt. Ltd., Noida, India for two years. He is now a Ph.D. student in the School of Mobile Computing and Communications of Jadavpur University.

**Subhadip Basu** received his B.E. degree in Computer Science and Engineering from Kuvempu University, Karnataka, India, in 1999. He received his Ph.D. (Engg.) degree thereafter from Jadavpur University (J.U.) in 2006. He joined J.U. as a senior lecturer in 2006. His areas of current research interest are OCR of handwritten text, gesture recognition, real-time image processing.

**MITA NASIPURI** received her B.E.Tel.E., M.E.Tel.E., and Ph.D. (Engg.) degrees from Jadavpur University, in 1979, 1981 and 1990, respectively. Prof. Nasipuri has been a faculty member of J.U since 1987. Her current research interest includes image processing, pattern recognition, and multimedia systems. She is a senior member of the IEEE, USA, Fellow of I.E (India) and W.A.S.T., Kolkata, India.